\documentclass[seceq]{jpsj2}
\usepackage{graphicx}
\usepackage{latexsym}
\usepackage{amsmath}
\usepackage{bm}
\usepackage{color}

\newcommand{\udarrow}[2]{\smash{\mathop{%
  \hbox to 0.6cm{$\rightleftharpoons$}}\limits^{#1}\limits_{#2}}}

\def\p{\partial}

\title{%
Selection of Crystal Chirality: Equilibrium or Nonequilibrium?
}

\author{%
Yukio \textsc{Saito}\thanks{yukio@rk.phys.keio.ac.jp}
 and Hiroyuki \textsc{Hyuga}\thanks{hyuga@rk.phys.keio.ac.jp}
}

\inst{%
Department of Physics, Keio University, Yokohama 223-8522 
}

\recdate{\today}

\abst{%
To study the solution growth of crystals composed of chiral organic molecules, 
a spin-one Ising lattice gas model is proposed.
The model turns out to be equivalent to the Blume-Emery-Griffiths model,
which shows an equilibrium chiral symmetry breaking at low temperatures.
The kinetic Monte Carlo simulation of crystal growth, however, demonstrates
that
 Ostwald ripening is a very slow process 
 with a characteristic time proportional 
 to the system size:
The dynamics is nonergodic.
It is then argued that by incorporating
 grinding dynamics, homochirality is achieved in a short time, 
independent of the system size.
Grinding limits cluster sizes to a certain range
 independent of system size and at the same time keeps
the supersaturation so high that population numbers of average-sized 
clusters grow. 
If numbers of clusters for two types of enantiomers differ by chance, 
the difference is amplified exponentially and the system rapidly approaches the homochiral state. 
Relaxation time to the final homochiral state is determined
by the average cluster size.
We conclude that the system should be driven and kept in a nonequilibrium state
to achieve homochirality.
}

\kword{%
homochirality, crystal growth, lattice gas model, grinding, 
kinetic Monte Carlo simulation, driven closed system
}

\begin{document}
\sloppy
\maketitle

\section{Introduction}

Spontaneous symmetry breaking is observed in various 
fields of physics.
In life, the chiral symmetry of organic molecules is broken,
and the origin of this one-handedness or homochirality 
has been a long mystery ever since 
 Pasteur found the molecular chirality\cite{pasteur48,japp98}.
The first theoretical model for explaining the spontaneous breaking 
of chiral symmetry in
life is proposed by Frank in terms of a set of rate equations 
 with linear autocatalysis and nonlinear mutual inhibition \cite{frank53}.
The actual chemical system that realizes the amplification of enantiomeric
excess (ee) was found by Soai and coworkers\cite{soai+95,sato+03}. 
The time development of the yield and ee of the products
was elucidated by the second-order
autocatalytic process, which provides nonlinearity.
A  theoretical model with an additional recycling process was found to achieve
homochirality \cite{saito+04}.
Recently, new chemical systems that
show ee amplification have been found \cite{mauksch+07}.

Chirality also appears in the structure and shape of crystals.
In fact, Pasteur found two enantiomers of the sodium ammonium salt of
tartaric acid, since each enantiomer formed separate crystals 
with mirror symmetric shapes and different enantiomers 
were resolved just by hand picking
\cite{pasteur48}.
When two enantiomers form separate crystals, 
the molecule is said to form a conglomerate. 
It is noted, however, that only 10 percent of chiral organic molecules 
form a conglomerate
and the rest crystallize into racemic crystals.

Intrinsic 
chirality is not a prerequisite for the formation of chiral crystals:
Achiral molecules can form chiral crystals.
The eminent example is quartz,
which can be either in the d- or l-form.
Sodium chlorate, NaClO$_3$, is another example that 
crystallizes into
a chiral form \cite{kipping+98}.
It forms a conglomerate, and d- and l-crystallites are 
separately formed.
However, by simply evaporating a sodium chlorate solution 
in a growth cell, 
almost the same amounts of d- and l-crystallites are grown,
 and the original solution ends up into a racemic mixture 
in the cell \cite{kipping+98,kondepudi+90}.
In 2005, Viedma discovered that when the solution is stirred violently
and the growing crystal is ground to small pieces, all crystals eventually 
turn into crystals of a single chirality in a short time\cite{viedma05}.
Later, Noorduin and coworkers\cite{noorduin+08a,noorduin+08b}
grew crystals of an organic molecule,
the imine of 2-methyl-benzaldehyde and phenylglycinamide, under grinding,
and they found that a molecular chirality converts to a single one.
Recently, other organic substances as amino acid are found to show 
similar deracemization phenomena.\cite{viedma+08,tsogoeva+09}

There have been many theoretical studies on the selection of crystal chirality
mainly based on rate equations\cite{uwaha04,saito+05,uwaha08}.
In our recent study \cite{saito+08}, we proposed 
a simple lattice gas model and performed Monte Carlo simulation
to compare results with those of  Noorduin's experiment.
Our simulation study revealed that the spontaneous chiral symmetry breaking 
occurs in a short time when one includes 
grinding and the nonlinear amplification of
chirality conversion on the 
surface of enantiomeric crystals.
\cite{saito+08}
In the present study, we analyze the equilibrium and nonequilibrium aspects
of the model in detail, and discuss what are the essential features of the
chiral symmetry breaking in crystal growth.

In the next section, a lattice model for chiral crystal growth, which has three
possible states on a lattice site, is presented. 
The model turns out to be equivalent to
the Blume-Emery-Griffiths model\cite{blume+71}
 with chiral symmetry breaking in equilibrium.
In \S 3, we describe kinetic processes consisting of
 crystal growth and dissolution, chirality conversion, and grinding.
Simulation results are described in \S 4. 
Without grinding, chirality is selected via the Ostwald ripening\cite{ostwald97} 
 of the two (or a few) largest clusters of different
enantiomers, and it takes an enormously long time to establish homochirality
with a relaxation time of the order of the system size: 
The system is nonergodic. By including the grinding procedure,
the relaxation time is shown to be independent of the system size, 
and the chirality is selected rapidly even in a large system. 
Discussions are given in \S 5.

\section{Equilibrium Selection of Chirality}
\setcounter{equation}{0}

We consider the crystal growth of organic molecules from a solution.
The organic molecules are chiral and
 can be in one of the two possible 
enantiomeric forms, classified as either R or S.
It is experimentally known that
the grown crystals can
show chiral symmetry breaking if they form a conglomerate, namely,
the R crystal grows separately from the S crystal.
Otherwise, if
chiral molecules form a racemic compound, both R and S enantiomers are
contained in a single crystallite and achieving homochirality is impossible.
This fact indicates that for the formation of a homochiral crystal,
bonding energies between a pair of 
the same kind of enantiomers, $J_{RR}$ and $J_{SS}$, should be larger than 
that between a heteropair, $J_{RS}$:
\begin{align}
J_{RR}=J_{SS}>J_{RS} .
\end{align}
Due to the microscopic chiral symmetry, bonding energies between the
same kind of enantiomers are assumed to be independent of the 
enantiomeric type.

We construct a lattice gas model such that
on each lattice site there is
either an R or S chiral enantiomer, or none.
The double occupancy of a lattice site is forbidden.
Because of simplicity, we consider a two-dimensional system where
the space is divided into a square lattice. 
The state of the system changes as molecules change their
sites or chirality.
When a pair of nearest-neighboring sites are occupied by molecules,
they form a bond with an energy gain.
In this section, we consider the equilibrium state of the system at a 
given temperature $T$. It is expected that 
at sufficiently low temperatures,
enantiomers of the same type will form clusters to gain energy,
and the chiral symmetry of the system may break spontaneously.

The model, in fact, reduces to a spin-one Ising model
studied by Blume, Emery and Griffiths\cite{blume+71},
which is called the BEG model.
The state on the site $i$ is described by the spin variable $S_i$ 
which, in our model, is assumed to take the value $+1$ if
the site is occupied by an R enantiomer, $-1$ if the site is occupied by an S enantiomer, and
0 if the site is vacant;
\begin{align}
S_i=
\begin{cases}
+1 & \mbox{occupied by R}\\
~0 & \mbox{vacant}\\
-1 & \mbox{occupied by S}.
\end{cases}
\end{align}
The total number of chiral molecules is given by the sum 
of $S_i^2$ as
\begin{align}
N=N_R+N_S=\sum_i S_i^2,
\end{align}
whereas the population difference of enantiomers is given by
the sum of $S_i$ as
\begin{align}
M=N_R-N_S=\sum_i S_i.
\end{align}
Here, $N_R$ and $N_S$ represent the numbers of R and S enantiomers, respectively,
and the summation runs over a square lattice of $L^2$ sites.
The average 
value of $S_i^2$ corresponds to the concentration of chiral molecules as
\begin{align}
c=\langle S_i^2 \rangle =\langle N \rangle/L^2,
\end{align}
and the average value of $S_i$ is related to the crystal enantiomeric excess 
(cee) or 
the chiral order parameter $\phi$ as
\begin{align}
c \phi =\langle S_i \rangle= \langle M \rangle/L^2 .
\end{align}

One can construct projection operators for each state as follows:
$\hat P_i(R)=S_i(S_i+1)/2$ is a projection operator for an R enantiomer since it
is unity if the site $i$ is occupied by
an R enantiomer, and zero otherwise. Similarly,
$\hat P_i(S)=S_i(S_i-1)/2$ and $\hat P_i(0)=1-S_i^2$ are projection operators
for an S enantiomer and a vacancy, respectively.
By the use of these projection operators, the interaction energy is written as
\begin{align}
E_0&=-\sum_{\langle i,j \rangle} [
J_{RR} \hat P_i(R)\hat P_j(R) +
J_{RS} (\hat P_i(R)\hat P_j(S) 
\nonumber \\
&\qquad \qquad + \hat P_i(S)\hat P_j(R)) +
J_{SS} \hat P_i(S)\hat P_j(S) ]
\nonumber \\
&=-\sum_{\langle i,j \rangle} (KS_i^2S_j^2+JS_iS_j)
\end{align}
with
\begin{align}
K=\frac{1}{2}(J_{RR}+J_{RS}) , \qquad J=\frac{1}{2}(J_{RR}-J_{RS}).
\end{align}
Here, the symmetry relation $J_{RR}=J_{SS}$ is utilized.
To discuss phase transitions, we introduce conjugate intensive 
variables, namely, chemical potentials $\mu_R$ and $\mu_S$, as
\begin{align}
&E=E_0-\mu_R N_R - \mu_S N_S
\nonumber \\
&=-\sum_{\langle i,j \rangle} (KS_i^2S_j^2+JS_iS_j) - \mu \sum_i S_i^2
-H \sum_i S_i
\end{align}
with
$\mu=(\mu_R+\mu_S)/2$ and $H=(\mu_R-\mu_S)/2$.
This is just the model discussed by Blume et al
\cite{blume+71}.
The field $H$ breaks the chiral symmetry externally. 
The spontaneous chiral symmetry breaking will take place
in the absence of an external field, or $H=0$.

For a macroscopic state defined by the numbers of R and S enantiomers, 
$N_R$ and $N_S$, respectively, 
the energy in  the mean field approximation is written as
\begin{align}
E=& -\frac{z}{2L^2} (J_{RR}N_R^2+J_{SS}N_S^2+2J_{RS}N_RN_S)
\nonumber \\
&-\mu (N_R+N_S) -H(N_R-N_S)
\end{align}
with the coordination number $z=4$ 
for a square lattice,
 and the entropy as
\begin{align}
S=k_B \ln \frac{L^2!}{N_R! N_S! (L^2-N_R-N_S)!} .
\end{align}
For simplicity, we take the Boltzmann constant to be unity, $k_B=1$, hereafter.
In terms of the molecular concentration
$c=(N_R+N_S)/{L^2}$
and the chiral order parameter $\phi=(N_R-N_S)/(N_R+N_S)$,
the free energy density $f=(E-TS)/L^2$ is expressed as
\begin{align}
f=& -(z/2)Kc^2-(z/2)J(c \phi)^2-\mu c - H c \phi
\nonumber \\
&+ T \Big[
c \, \frac{1+\phi}{2} \ln \Big( c \, \frac{1+\phi}{2}\Big) 
+c \, \frac{1-\phi}{2} \ln \Big( c \, \frac{1-\phi}{2} \Big)
\nonumber \\
&+(1-c)\ln(1-c)
\Big] .
\end{align}
By minimizing $f$ in terms of $c$ and $\phi$ at the fixed intensive parameters
$\mu$ and $H$, one obtains the self-consistent equations 
for equilibrium values of $c$ and $\phi$  as
\begin{align}
\phi&=\tanh \frac{zJc \phi+H}{T},
\nonumber \\
c&=\frac{2 \cosh[(zJc \phi+H)/T]}{\exp[-(zKc+\mu)/T]
+2 \cosh[(zJc \phi+H)/T]}
.
\end{align}
To study the spontaneous chiral symmetry breaking, the external field
$H$ is set to be zero, whereas the chemical potential $\mu$ is chosen
appropriately to fix the concentration $c$ at a desired value.

\begin{figure}[tbh]
\begin{center} 
\begin{minipage}{0.48\linewidth}
\begin{center}
\includegraphics[width=1.0\linewidth]{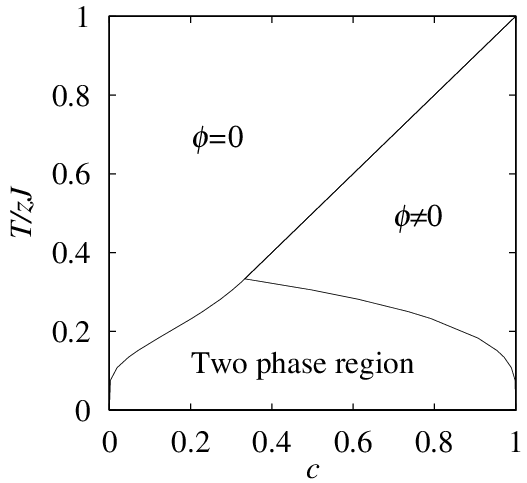}\\
(a)\\
\end{center}
\end{minipage}
\begin{minipage}{0.48\linewidth}
\begin{center}
\includegraphics[width=1.0\linewidth]{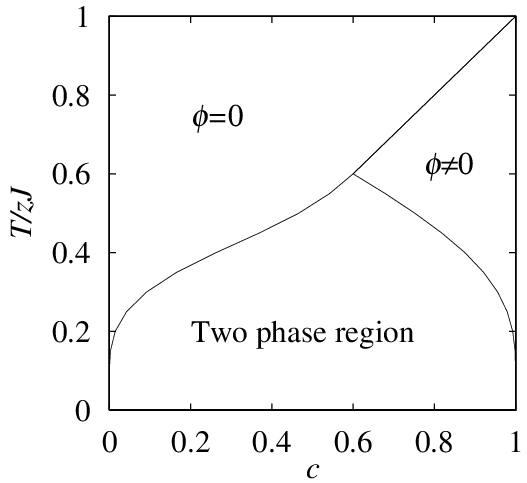}\\
(b)\\
\end{center}
\end{minipage}
\end{center} 
\caption{Equilibrium phase diagram of BEG model.
(a) $K=0$. (b) $K=J$.
}
\label{fig1}
\end{figure}

For $H=0$, there are two solutions; one is racemic with $\phi=0$, and
the other is chiral with $\phi \ne 0$. The true equilibrium state is 
that with the minimum free energy $f$.
For the case $K=0$ where $J_{RS}=-J_{RR}<0$ so that R and S are repulsive, 
the phase diagram is shown in Fig.1(a), which is essentially the same
as that depicted by Blume et al
\cite{blume+71}.
(Note that the He$^3$ concentration $x$ 
in ref. 19 is related to 
our molecular concentration $c$ by the relation $x=1-c$.)
At high temperatures, $T>zJ$, the system is racemic for all values of
concentration $c$. 
Below the second-order transition temperature $zJc$, 
the chiral symmetry is broken with a finite value of $\phi$ at a given $c$.
Below the tricritical temperature $T_3=zJ/3$, 
the system separates into the following two phases:
a low-concentration phase that is racemic and a high-concentration phase
that is chiral and has a finite $\phi$.
High- and low-concentration phases correspond to 
an aggregate of chiral crystals and a dilute solution, respectively.

In the following kinetic Monte Carlo simulation, we consider the case
with a vanishing bond energy between 
different enantiomers,
R and S; $J_{RS}=0$. 
It corresponds to the interaction
parameters $K=J=J_{RR}$. 
The phase diagram in the mean-field approximation
turns out to be essentially the same as the one at $K=0$,
 as is shown in Fig. 1(b).
 The tricritical point is now $T_3=3zJ/5$.
In the following simulation, the concentration of chiral molecules 
is fixed to be as low as $c=0.1$. The phase separation in the mean-field
approximation takes place at 
$T/zJ=0.3066$ in this approximation, as is seen in Fig. 1(b).
However, in low-dimensional systems, and 
for a finite-sized system in particular,
 thermal fluctuation is,
in general,  large, suppressing ordering. 
 Therefore, in the following kinetic Monte Carlo simulation,
a sufficiently low temperature is chosen to ensure that the system 
phase-separates with a broken crystal chirality.

\section{Kinetic Monte Carlo Simulation}
\setcounter{equation}{0}

In a previous section, it is shown that the chiral symmetry breaking 
takes place as an equilibrium phase transition for a conglomerate system.
The equilibrium argument, however, cannot tell us about the dynamic process,
for example, how long we have to wait for the equilibrium
state to be achieved.
In the discussion on the equilibrium state, 
the chirality and position of a molecule are assumed to change instantaneously.
In the actual experiment of organic molecules, only monomers can change their
chirality in a solution \cite{noorduin+08a}.
Molecules incorporated into a cluster have to be dissolved again in a solution
before they can change their chirality.
Thus, it takes a long time to achieve homochirality.
For example, in the final equilibrium state with homochirality at $T=0$,
all molecules coagulate into a single large crystal
with a single chirality. To reach this final state,
crystals have to compete with each other and coarsen. 
This coarsening process, usually called
Ostwald ripening \cite{ostwald97}, proceeds very slowly and
thus makes the realization of chirality selection practically impossible
in an observable time. To circumvent this obstacle, 
grinding plays the essential role
\cite{viedma05,noorduin+08a}.
A dynamic simulation model with grinding was proposed by us previously
\cite{saito+08}.
We report here the effect of grinding on the temporal evolution of 
chirality in detail.
As a measure for characterizing the dynamics of chiral symmetry breaking,
we adopt a saturation time $t_s$ for a system that starts from a racemic
initial state to reach the final homochiral state.
We study its dependence on the total number of molecules $N$.

Dynamics of our lattice gas model consists of three processes:
(1) crystal growth and dissolution, (2)  chirality conversion, and 
(3) periodic grinding.

\subsection{Crystal growth and dissolution}

When a chiral molecule is isolated in a square lattice, 
it can hop to an arbitrary empty site at a rate $k$. 
This long-distance jump reflects the fact that
the solution is strongly stirred 
and diffusion to neighboring sites 
has negligible effect in experiments.
When the molecule lands on a site with some neighbors of the same chirality,
bonds are formed. 
When there are $n$ nearest neighbors of the same enatiomeric type,
$n$ homobonds are formed with 
the local energy gain $-nJ$. These bonds have to be broken
for the next jump, and accordingly, the jump rate decreases to
\begin{align}
ke^{-nJ/T} .
\label{eq3.1}
\end{align}
Since the highly coordinated molecule dissolves less frequently, 
homoclusters remain stable and grow into large crystals.
In the simulation, highly coordinated molecules with $n=$3 and 4 
are not dissolved at all.
The rate $\exp(-J/T)$ is chosen to be $10^{-3}$,
which corresponds to $T/J=(3 \ln 10)^{-1}=0.1448$. 
For the simulation at a concentration $c=0.1$,
this temperature $T/zJ=0.1448/4=0.036$ is very low compared with
the expected phase-separation line $T/zJ \approx 0.31$ in the mean-field 
calculation, shown in the phase diagram in Fig.1 (b).
At this temperature,
the system is expected to separate into a very dilute solution
and an almost homochiral crystal (with $c \approx 1$) 
in the mean-field approximation.

\subsection{Chirality conversion}

Experimentally, it is known that a monomer changes chirality
in the solution, if an appropriate agent is added \cite{noorduin+08a}.
This spontaneous conversion of molecular chirality is an
 essential process for chirality selection,
and we denote the conversion rate as $\nu_0$; 
\begin{align}
\nu_0: R \rightleftharpoons S
\label{eq3.15}
\end{align}
when the monomers are isolated.
In the following simulations,
time is measured in units of $\nu_0$ and the jump rate of an isolated
 monomer is assumed to be as large as  $k=10^4 \nu_0$, since the
solution is vigorously stirred with a large jump rate.
In a previous study, we assumed an enhancement of the chirality conversion rate
to $\nu_n$, when a monomer is in contact with $n$ opposite enantiomers.
The present study reveals that such an enhancement turns out to be inessential, 
as is discussed in the next section.

\subsection{Periodic grinding}

\begin{figure}[tbh]
\begin{center} 
\begin{minipage}{0.6\linewidth}
\begin{center}
\includegraphics[width=1.0\linewidth]{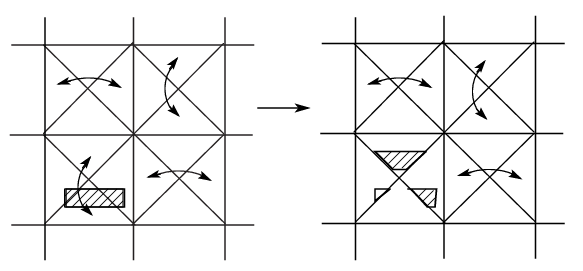}\\
(a)\\
\end{center}
\end{minipage}
\hspace{0.05\linewidth}
\begin{minipage}{0.28\linewidth}
\begin{center}
\includegraphics[width=1.0\linewidth]{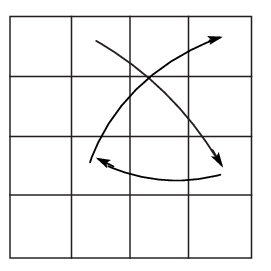}\\
(b)\\
\end{center}
\end{minipage}
\end{center} 
\caption{Grinding 
consists of (a) diagonal cut followed by up-down or 
left-right triangle exchange in each square mesh
and (b) random shuffle 
of all meshes.
}
\label{fig2}
\end{figure}

In aforementioned experiments\cite{viedma05,noorduin+08a},
 many glass balls are put 
in the crystal growth cell,
and when magnetic stirrers rotate to stir the solution, glass balls
rotate and collide randomly with growing crystallites
and smash them into small fragments.
We simulate this grinding process by the following procedures:@
First, the 
whole
system is shifted randomly in space and then
divided into square mesh cells of side length $\ell$.
Each mesh is cross-cut by two diagonals into four triangles, and
up-down or left-right triangles are exchanged randomly (see Fig. 2(a)).
If diagonal lines cut through faceted crystal clusters, 
they break crystal facets and create small fragments covered by
high-index faces. 
Then, all meshes are interchanged randomly (Fig. 2(b)).
Clusters with a linear dimension larger than about $\ell/2$ 
would be broken, 
although it is still possible that 
some clusters in neighboring meshes may reconnect to form large ones after grinding.
Our interest is focused on the effect of grinding on the development of homochirality
 as the system size $L$ is varied.

The grinding process is introduced periodically in the simulation 
with a frequency $f$ or
after every interval of time, $f^{-1}$.
The relation between the grinding frequency $f$ and the chirality selection is 
also an interesting aspect.

\section{Simulation Results}
\setcounter{equation}{0}

The most interesting point we want to understand is 
the temporal development of chiral symmetry breaking.
Therefore, in an initial state, equal numbers of R and S enantiomers 
are distributed randomly in the system; 
$N_R=N_S=N/2$, i.e., the initial system is racemic, $\phi(t=0)=0$.
During a simulation run, one enantiomer starts to dominate by chance
 and $\phi$ takes a nonzero value.
The system eventually reaches a homochiral state with $|\phi|=1$
at a saturation time $t_s$. 
We expect that this saturation time $t_s$ will characterize the dynamics 
of the chirality selection.
The time $t_s$ depends on many parameters, but our focus here is its system size dependence.
For its study, the linear size of the system, $L$, is varied, keeping the
chiral molecule concentration $c=N/L^2$ at a fixed value $c=0.1$.

\subsection{Ostwald ripening}

When the simulation is performed with neither grinding 
nor an enhancement of chirality conversion rate, 
molecules form two large clusters of different enantiomers rather
rapidly in our small system, and then, the competition begins between them. 
For example, in one MC simulation run of a system with  a
size $L=160$, there are four clusters,
two R's and two S's, at a time $t \nu_0 \approx 2.5 \times 10^4$ (Fig. 3(a)),
and the cluster number is reduced to two with one R and one S each at 
$t \nu_0 \approx 8 \times 10^4$ (Fig. 3(b)).
Then, the competition of chirality conversion starts.
One enantiomer dominates slightly by chance and eats up the
opposite very slowly. Eventually, only a single crystal cluster remains
after a saturation time $t_s$.
In the example,
$t_s \nu_0\approx 1.75 \times 10^6$, as shown in Fig. 3(c).

\begin{figure}[htb]
\begin{center} 
\begin{minipage}{0.30\linewidth}
\begin{center}
\includegraphics[width=1.0\linewidth]{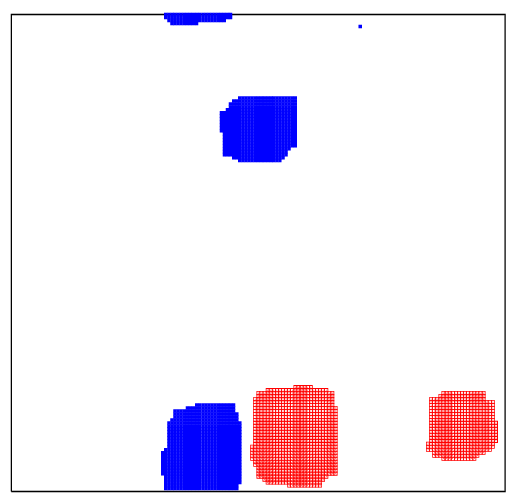}\\
(a)\\
\end{center}
\end{minipage}
\begin{minipage}{0.30\linewidth}
\begin{center}
\includegraphics[width=1.0\linewidth]{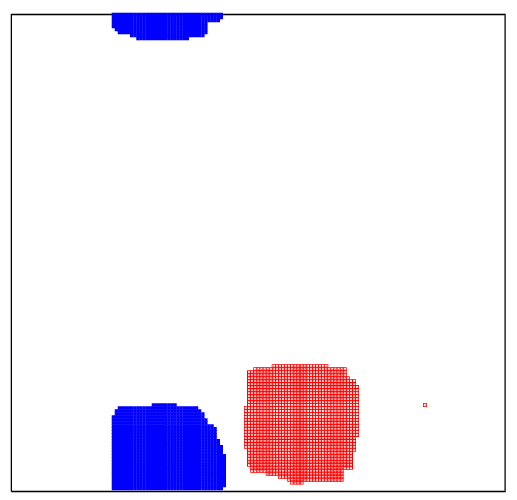}\\
(b)\\
\end{center}
\end{minipage}
\begin{minipage}{0.30\linewidth}
\begin{center}
\includegraphics[width=1.0\linewidth]{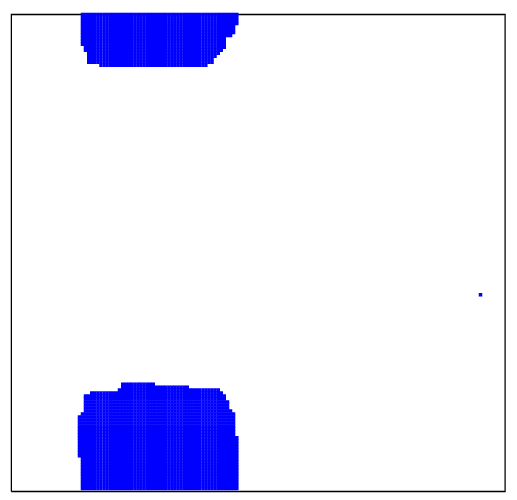}\\
(c)\\
\end{center}
\end{minipage}
\end{center} 
\caption{(Color online) Ostwald ripening: 
There are (a) four clusters at
$t\nu_0=25,000$, (b)  
two clusters at
$t\nu_0=80,000$, 
and (c) a single cluster at $t_s\nu_0=1,750,000$. 
The system size is $L^2=160^2$ at a number of molecules $N=2560$ 
at a 
concentration 
$c=0.1$.
}
\label{fig3}
\end{figure}

What really matters is the fact that
the time $t_s$ depends on the system size, as shown by open circles in Fig. 4.
The size dependence is approximately proportional to the number
$N$ of chiral monomers in the system. In Fig. 4, the line 
$t_s\nu_0 =500N$ is drawn as a guide to the eyes.
The size dependence of $t_s$ indicates that for a macroscopic system, 
the homochiral state cannot be achieved in a practical time, 
even though the equilibrium argument
assures the homochirality of the conglomerate system. The system is nonergodic.
\begin{figure}[thb]
\begin{center} 
\includegraphics[width=0.7\linewidth]{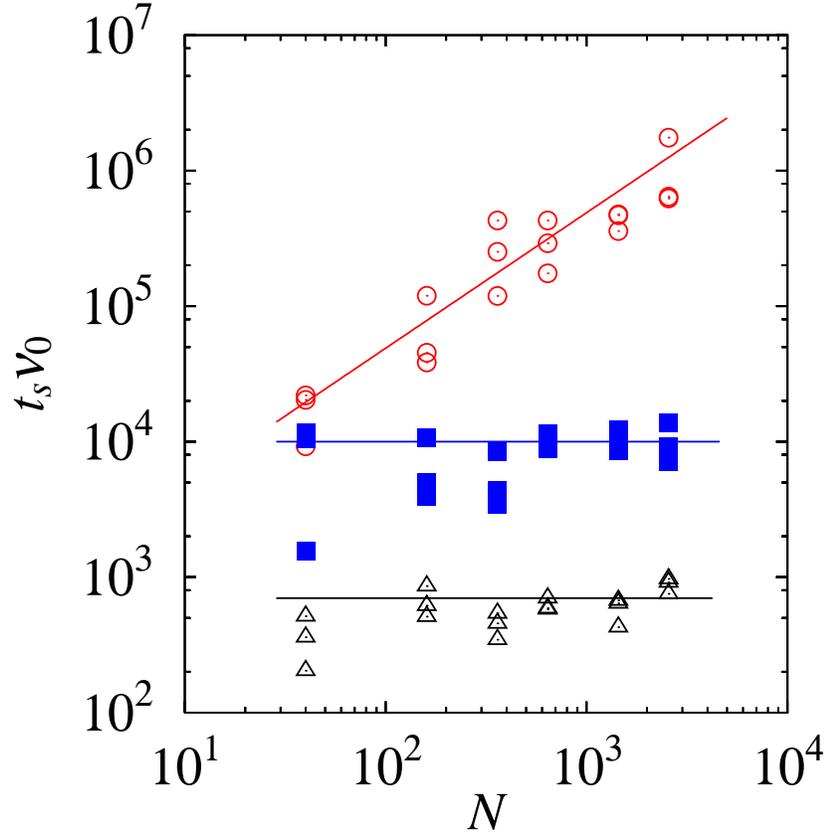}
\end{center} 
\caption{(Color online) Log-log plot of the saturation time $t_s$ versus
total molecular number $N$ 
(a) without grinding (open circle), (b) with grinding but
without 
enhancement of chirality conversion rate 
(filled box), and 
(c) with both grinding and rate enhancement (open up-triangle).
}
\label{fig4}
\end{figure}

\begin{figure}[thb]
\begin{center} 
\includegraphics[width=0.7\linewidth]{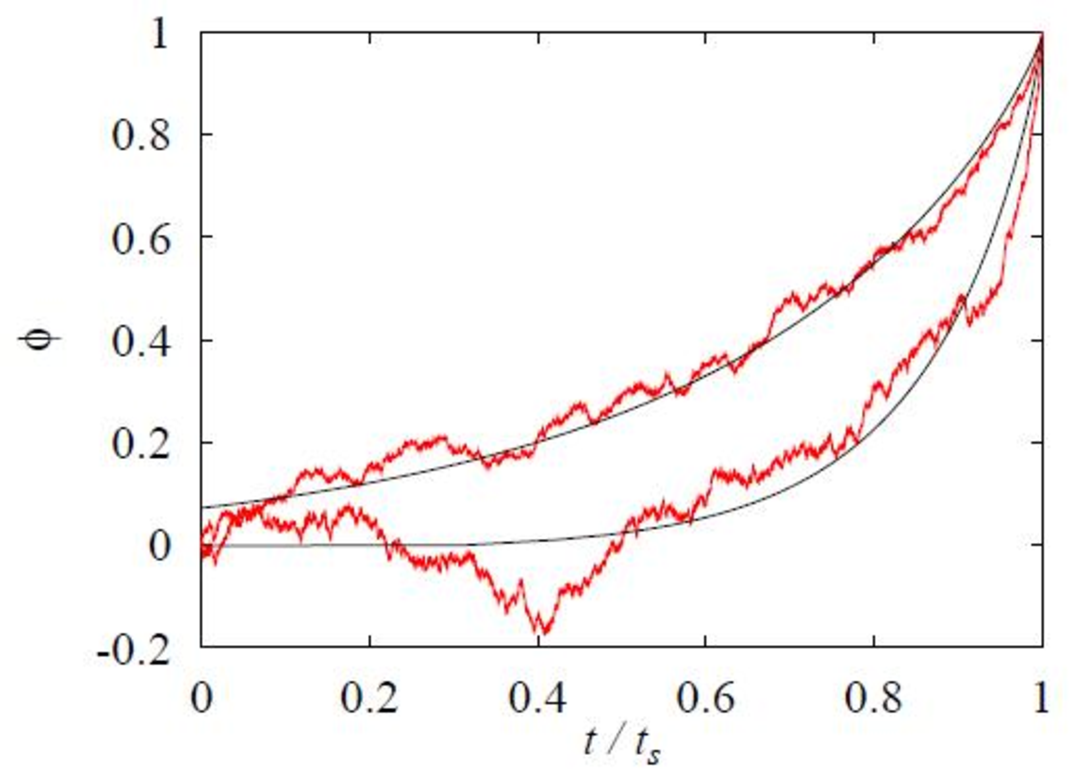}
\end{center} 
\caption{(Color online) Temporal evolution of $\phi$ for Ostwald ripening.
Two simulation results for a 
system
 with a size $L^2=160^2$ and $N=2560$
are compared with theoretical curves, eq. (\ref{eq4.19}),
 with an appropriate 
 choice
  of $C$.
}
\label{fig5}
\end{figure}

Here, we would like to present a qualitative argument about the time 
development of chiral order
parameter
$\phi$ after each enantiomer forms only a large single cluster of its kind,
as in Fig. 3(b).
Let there be $N_R$ molecules of R-type enantiomers and $N_S$ molecules of 
S-type enantiomers. Then, linear dimensions of clusters are approximately 
$\ell_R \approx \sqrt{N_R}$ and $\ell_S \approx \sqrt{N_S}$.
The time evolution of the cluster size may be written as
\begin{align}
\frac{d \ell_R}{dt}= K \Big(\Delta \mu - \frac{\sigma}{\ell_R} \Big),
\nonumber \\
\frac{d \ell_S}{dt}= K \Big(\Delta \mu - \frac{\sigma}{\ell_S} \Big),
\label{eq4.16}
\end{align}
with a certain appropriate kinetic coefficient $K$ since the growth is controlled by
surface kinetics \cite{saito96}.
The driving force $\Delta \mu$ is related to the concentration of enantiomeric
monomers in the solution. Since the Ostwald ripening takes place very slowly, 
enantiomers change their chirality frequently
 enough so that one can assume that
$\Delta \mu$ is the same for both chiral species, R and S. 
$\sigma$ represents the effect of surface tension.
From the above eq. (\ref{eq4.16}), one can derive the evolution equation 
for the size difference 
\begin{align}
\frac{d (\ell_R- \ell_S) }{dt}= 
K \sigma (\ell_S^{-1} - \ell_R^{-1} ).
\end{align}
Since $\ell_R^2=N_R=N(1+\phi)/2$ and $\ell_S^2=N_S=N(1-\phi)/2$, 
the chiral order parameter $\phi$ evolves as
\begin{align}
\frac{d \phi}{dt}= 
\frac{4 K \sigma}{N} \frac{1-\sqrt{1-\phi^2}}{\phi} .
\end{align}
It is solved analytically as
\begin{align}
\sqrt{1-\phi^2}+\ln (1-\sqrt{1-\phi^2})=
(4 K \sigma/N) t+C
\label{eq4.19}
\end{align}
with an integration constant $C$ to be determined by the intial condition.
Two examples of temporal evolution of the chiral order 
parameter
$\phi$ 
are shown in Fig. 5.
The system has a size $L^2=160^2$ with a concentration $c=0.1$, and
an elapsed time is normalized by the saturation time $t_s$. 
The saturation time $t_s$ differs by a factor of three for two runs,
depending on the initial incubation time.
However, simulation results are well fitted to the theoretical behavior 
in eq. (\ref{eq4.19}), drawn by smooth curves,
with appropriately chosen values of constant $C$.

By assuming a small initial cee $|\phi(0)| \ll 1$,
the order parameter develops as $d \phi/dt =2K \sigma \phi/N$ 
and increases exponentially as
\begin{align}
\phi(t) = \phi(0) \exp(2K \sigma t/N) .
\end{align}
It then reaches the saturation value $|\phi(t_s)|=1$ at a time given by
\begin{align}
t_s=\frac{N}{2K \sigma} \ln |\phi(0)|^{-1},
\end{align}
indicating that the saturation time $t_s$ is proportional to the cluster size
$N$, as observed in Fig. 4.
By assuming an initial fluctuation of the order $|\phi(0)| \approx \sqrt{1/N}$,
$t_s$ may have a weak logarithmic size dependence; 
$t_s \approx N \ln N/4K \sigma$, although it is difficult to confirm this 
dependence in numerical simulations.

\subsection{With grinding}

We introduce a grinding procedure with the frequency $f/\nu_0=1$.
Then, clusters are broken into pieces after every time interval $1/f$
and competition among small clusters repeats:
All throughout the simulation, the system contains many 
small clusters, as shown in Fig. 6.
In a racemic state, there are as many R clusters as S clusters,
as shown in Fig. 6(a).
In a chiral state, the numbers of R and S clusters differ,
though cluster sizes are distributed similarly.
In the homochiral state shown in Fig. 6(b),
all the clusters belong to one enantiomer, but the size distribution
appears the same as that in Fig. 6(a).

\begin{figure}[htb]
\begin{center} 
\begin{minipage}{0.48\linewidth}
\begin{center}
\includegraphics[width=1.0\linewidth]{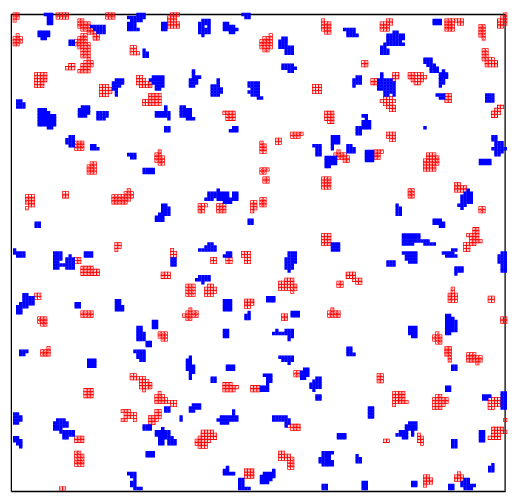}\\
(a)\\
\end{center}
\end{minipage}
\begin{minipage}{0.48\linewidth}
\begin{center}
\includegraphics[width=1.0\linewidth]{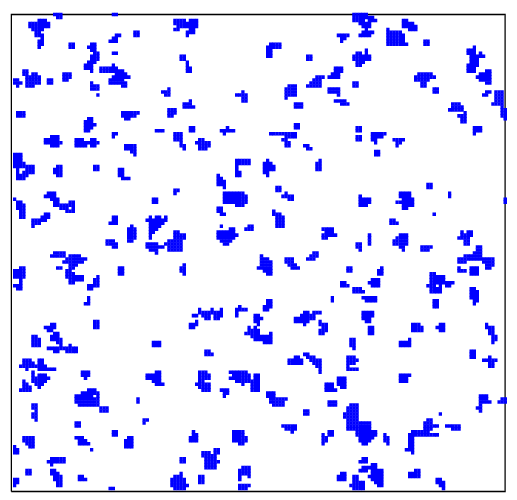}\\
(b)\\
\end{center}
\end{minipage}
\end{center} 
\caption{(Color online) Cluster configuration with grinding  
(a) at the 
intermediate
 time $t\nu_0=1,800$, and (b) at the 
 saturation
 time of
homochirality $t_s\nu_0=7,200$. 
The system size is $L^2=160^2$ at a number of molecules $N=2560$ at a
 concentration $c=0.1$.
}
\label{fig6}
\end{figure}

\begin{figure*}[thb]
\begin{center} 
\begin{minipage}{0.32\linewidth}
\begin{center}
\includegraphics[width=1.0\linewidth]{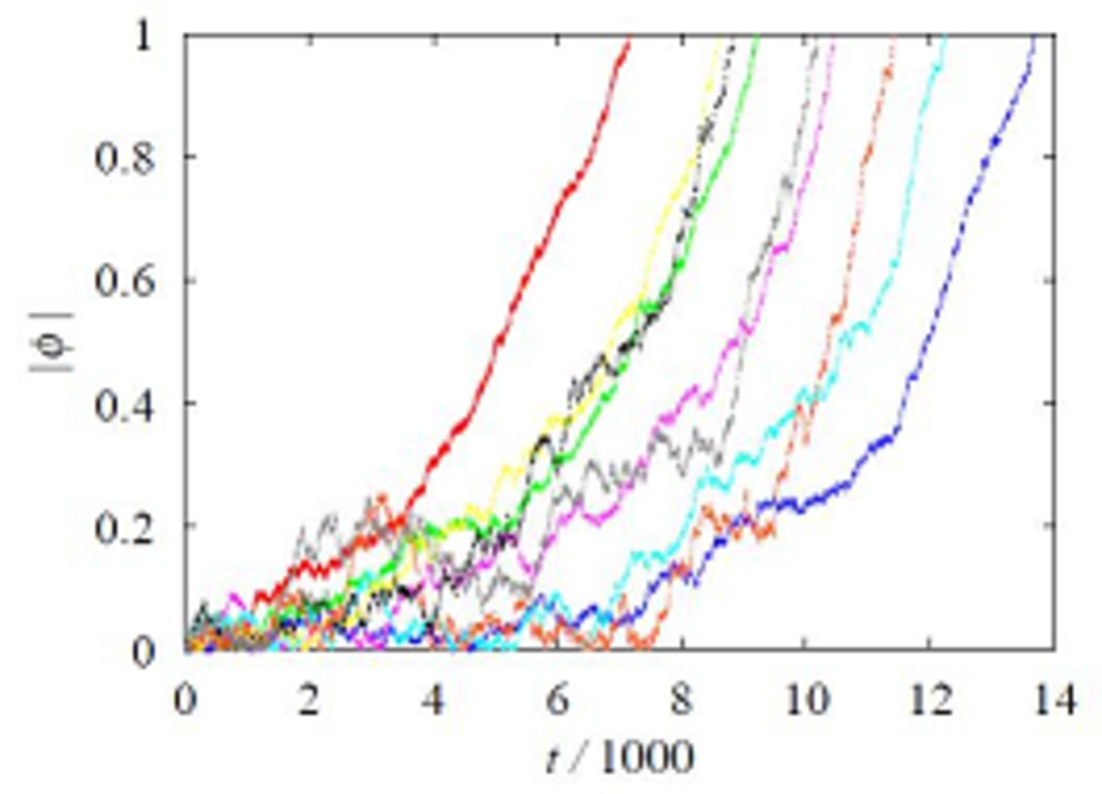}\\
(a)\\
\end{center}
\end{minipage}
\begin{minipage}{0.32\linewidth}
\begin{center}
\includegraphics[width=1.0\linewidth]{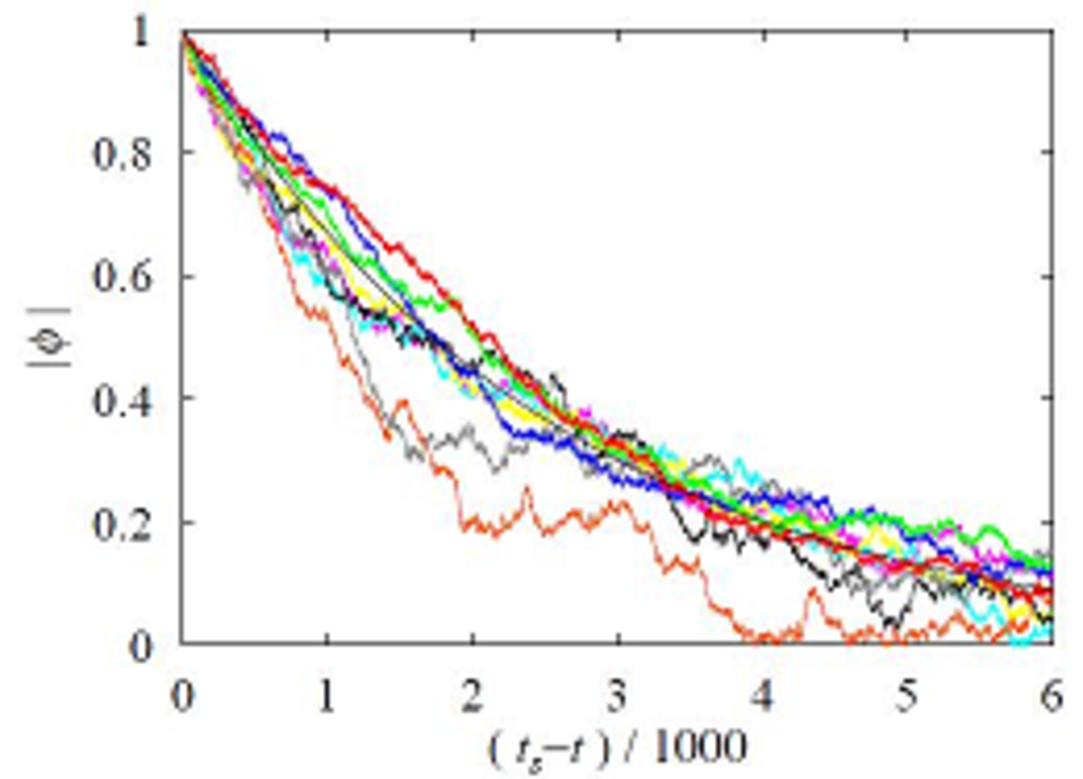}\\
(b)\\
\end{center}
\end{minipage}
\begin{minipage}{0.32\linewidth}
\begin{center}
\includegraphics[width=1.0\linewidth]{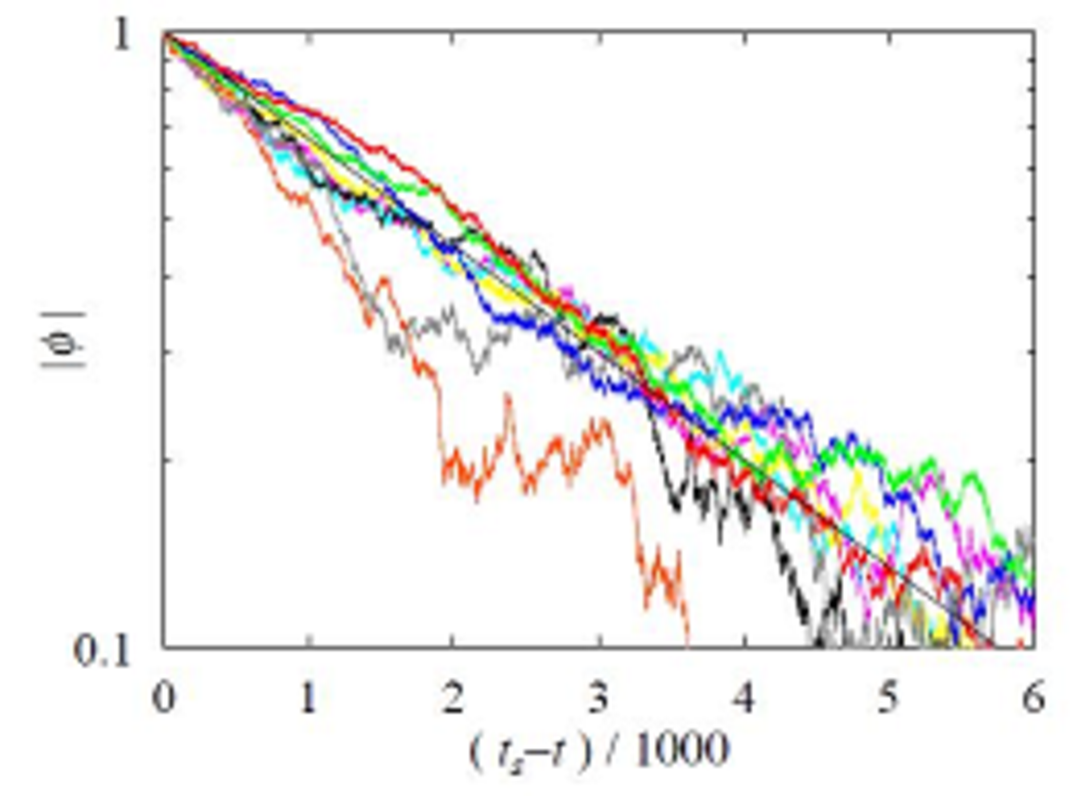}\\
(c)\\
\end{center}
\end{minipage}
\end{center} 
\caption{(Color online) (a) Temporal evolution of cee, $|\phi|$, 
for systems of three sizes: $L=80$, 120 and 160. 
There are three samples for each system size.
No systematic size dependence is observed for the initial nucleation period.
(b) $|\phi|$ close to the saturation, plotted versus $t_s-t$,
and (c) $|\phi|$ in a semilogarithmic plot.
The exponential behavior $\exp(-(t_s-t)/2500)$ is drawn in (b) and (c)
as a guide to the eyes.
}
\label{fig7}
\end{figure*}

The chiral order parameter $\phi$ increases steadily, as shown in Fig. 7(a).
It  reaches homochirality $|\phi|=1$ rapidly in comparison
 with the case without grinding, though
the saturation time $t_s$ is distributed (Fig. 7(a)).
$t_s$ does not show any systematic dependence on the
system size $L$ or the total molecular number $N$, 
as shown by filled boxes in Fig. 4.
For a large system, the saturation time $t_s$ is enormously shorter
than that in the case without grinding,
and the homochirality should be established rapidly.
We think that this explains essentially  the experimental achievement
of homochirality in a short time, observed by
Viedma \cite{viedma05} and Noorduin and coworkers\cite{noorduin+08a,noorduin+08b}.

The scattering of the saturation time $t_s$ is due to large fluctuations of
incubation time
before the chiral order parameter $\phi$ deviates from the
initial racemic state appreciably.
Incubation time is determined stochastically, and thus, the initial
start-up time for cee $|\phi|$ to shoot up depends on the simulation setup, 
more precisely stated,
on the random number sequence, as shown in Fig. 7(a).
Once cee takes a certain finite value beyond the thermal fluctuation, it
approaches the final homochiral state in a universal manner, 
as shown in Fig. 7(b).
In fact, the semilogarithmic plot in Fig. 7(c) indicates that $|\phi|$ increases
exponentially with time.
We want to understand this temporal behavior.

Without grinding, homochirality is achieved by a single large cluster
that wins the competitive process of Ostwald ripening. 
Under grinding with a finite mesh size $\ell$, 
as described in the previous section,
large clusters are broken into pieces and many clusters survive
whose size $s$ is distributed in a limited range, as evident in Fig. 6. 
From the number $C_{R,S}(s)$ of clusters  of size $s$ of each enantiomer, 
one obtains the average cluster size as
\begin{align}
\bar s_{R,S} = \frac{\sum_s s^2C_{R,S}(s)}{\sum_s s C_{R,S}(s)} ,
\end{align}
where the denominator represents the total number of enantiomeric
molecules
\begin{align}
N_{R,S}= \sum_s s C_{R,S}(s) .
\end{align}
We define the total number of clusters of each enantiomeric type by
\begin{align}
C_{R,S}= \sum_s  C_{R,S}(s) .
\label{eq4.24}
\end{align}

\begin{figure*}[hbt]
\begin{center} 
\begin{minipage}{0.48\linewidth}
\begin{center}
\includegraphics[width=1.0\linewidth]{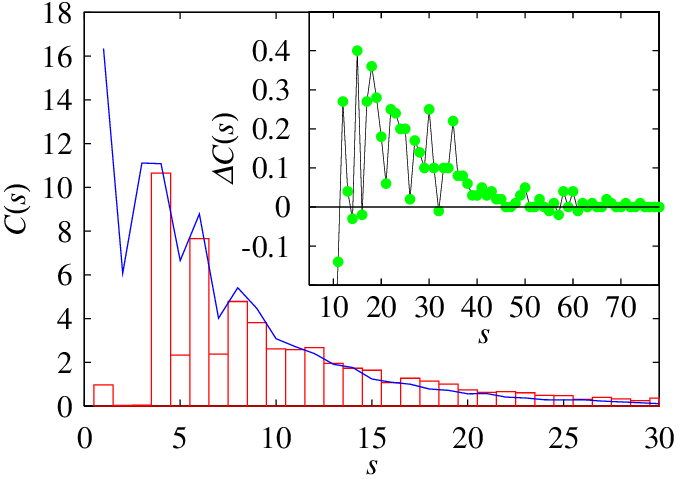}\\
(a)\\
\end{center}
\end{minipage}
\begin{minipage}{0.48\linewidth}
\begin{center}
\includegraphics[width=1.0\linewidth]{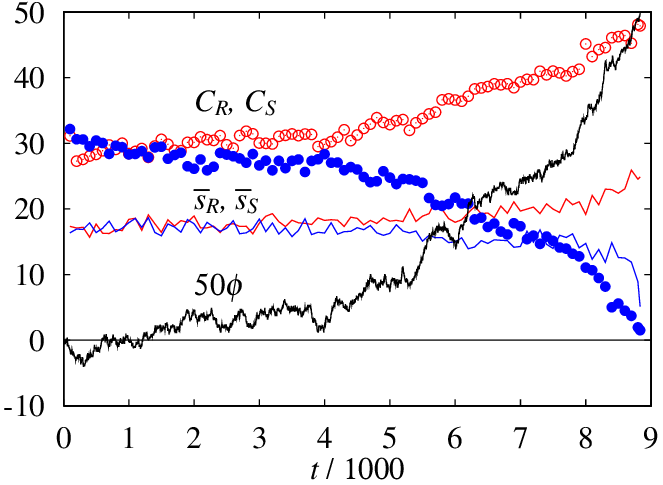}\\
(b)\\
\end{center}
\end{minipage}
\end{center} 
\caption{(Color online) (a)  
Cluster size distribution under grinding 
without enhancement of chirality conversion rate on the cluster.
Boxes represent distribution just before grinding $C_b(s)$, 
and lines just after grinding $C_a(s)$.
Inset shows the difference between the two $\Delta C(s)=C_b(s)-C_a(s)$,
which is negative for small $s$ (not shown) and positive for large $s$. 
The system size is $L^2=80^2$ at a number of molecules $N=640$ 
at a concentration $c=0.1$.
(b) Temporal evolution of average cluster sizes $\bar s_R$, $\bar s_S$ and
total cluster numbers $C_R$ and $C_S$ for two enantiomeric types.
The chiral order parameter $\phi$ is plotted as well.
}
\label{fig8}
\end{figure*}

An example of the
total cluster size distribution $C(s)=C_R(s)+C_S(s)$ is shown in Fig. 8(a).
It is obtained in a simulation run of a system with a size $L^2=80^2$ 
containing $N=640$ molecules at a time $t\nu_0=3000$, averaged
over an interval of time $\Delta t\nu_0=100$.  
The grinding mesh size is set at $\ell=20$.
Just after the grinding, many small clusters are produced as shown by
a curve 
$C_a(s)$ 
in Fig. 8(a), and the supersaturation is high.
Then, large clusters 
 grow by incorporating monomers.
As time proceeds, supersaturation decreases and
small clusters dissolve further to feed large clusters. 
As observed in the size distribution 
$C_b(s)$ 
after one period of
interval $1/f$, just before the next grinding,
the number of monomers decreases and small clusters disappear, 
whereas the number of large
clusters increases, as shown by a box graph in Fig. 8(a). 
The inset of Fig. 8(a) shows that the difference $\Delta C(s)=C_b(s)-C_a(s)$ 
is mostly positive for large sizes, indicating that large clusters grow 
to coarsen in this period.
There remains exceptionally many clusters 
with sizes 4 and 6, 
since they are compact in shape and 
energetically stable.

At the time $t \nu_0=3000$ shown in Fig. 8(a), 
there is not much difference in 
average cluster size between the R and S enantiomers, $\bar s_R \approx 18.9$ and 
$\bar s_S  \approx 16.6$, as well as in respective total cluster number,
$C_R \approx 30.2$ and $C_S  \approx 27.6$.
In later time, they differ, as shown in the
temporal evolution in Fig. 8(b).
In this figure, the average cluster sizes $\bar s_R $ and $\bar s_S $ 
are shown by curves
 and the cluster populations $C_R$ and $C_S$ are shown
by symbols.
The time evolution of the chiral order parameter $\phi$ is also depicted
in Fig. 8(b), and the homochiral state is achieved at $t_s \nu_0=8834$. 
It is evident that as $|\phi|$ increases,
cluster size and population start to differ for two species.
The average cluster size, however, does not vary markedly until the very end of
chirality selection.
The size is in any case distributed rather wide, as is apparent in Fig. 8(a).
The average size is mainly controlled by the mesh scale $\ell$
and grinding frequency $f$. 
It is found to be independent of the system size $L^2$.
On the contrary, the cluster population increases proportionally to $L^2$.
Also, it starts to differ systematically in an early stage,
even though the total cluster population $C=C_R+C_S$ 
remains stationary, around 58 in Fig. 8(b), until shortly before $t_s$.
This indicates that the numbers of clusters of two enantiomers compete
during the chirality selection.
We therefore consider the selection process in terms of
 cluster populations, $C_R$ and $C_S$, regarding a constant average size,
 $\bar s_R=\bar s_S=\bar s$.

Grinding splits large clusters to small ones, but it does not change
the chirality of molecules. Chirality variation occurs only 
during the coarsening period between two successive grinding procedures. 
Therefore, let us consider the evolution of the size distribution 
$\{C_R(s,t),\: C_S(s,t);\: s=1,2, \cdots \}$ after the $i$th grinding
until the next one, or at an interval of time, 
$i/f < t < (i+1)/f$.
Evolution is generally described in terms of the Becker-Doering equation
\cite{becker+35,saito96} as
\begin{align}
\frac{\p C_R(s,t)}{\p t}=&\sum_{s'} [W(s' \rightarrow s) C_R(s',t)
-W(s \rightarrow s') C_R(s,t)] 
\nonumber \\
&+\delta_{s,1} \nu_0[C_S(1,t)-C_R(1,t)],
\end{align}
where a term with Kronecker's delta $\delta_{s,1}$ represents the chirality
conversion process for the monomer ($s=1$).
$W$ generally represents the transition probability in size space.
Then, the total number of R enantiomers varies as
\begin{align}
&\frac{d N_R(t)}{d t}=\frac{\p }{\p t}(\sum_s sC_R(s,t))
\nonumber \\
&=\sum_{s} v(s) C_R(s,t)
+ \nu_0[C_S(1,t)-C_R(1,t)]
\end{align}
with the "velocity" in size space 
\begin{align}
v(s)=\sum_{s'} (s'-s)W(s \rightarrow s') .
\end{align}
Since the total number of molecules $N=N_R+N_S$ is conserved
as is expressed as
\begin{align}
\frac{d N}{d t}=\sum_{s} v(s) [C_R(s,t)+C_S(s,t)] =0,
\end{align}
$v(s)$ should change sign as $s$ varies. In fact, there is a
critical size $s_c$, and $v(s)$ is negative
for smaller sizes, $s<s_c$, and positive for $s>s_c$,
in accordance with the sign change of
$\Delta C(s)$ shown in the inset of Fig. 8(a).
Those clusters with negative $v(s)$, in fact, disappear during this coarsening
period, as shown in Fig. 8(a). 
Schematically, velocity may be represented as
\begin{align}
v(s)=K\sqrt{s} ( \Delta \mu - \sigma s^{-1/2}) ,
\end{align}
with $s_c=(\sigma/\Delta \mu)^2$.
The supersaturation $\Delta \mu$, in fact, depends on the instantaneous
monomer density.

For the difference between the total numbers of two enantiomers, 
$M=N_R-N_S$, we have 
\begin{align}
\frac{d M}{d t}=&\sum_{s} v(s) [ C_R(s,t)- C_S(s,t)]
\nonumber \\
&+ 2\nu_0[C_S(1,t)-C_R(1,t)].
\label{eq4.30}
\end{align}
The racemic state specified by
\begin{align}
N_R=N_S=\frac{N}{2}, \quad  C_R(s,t)= C_S(s,t)
\end{align}
 is a stationary solution of eq.(\ref{eq4.30}).
If the total numbers of two enantiomers differ, such that $M=N_R-N_S>0$ 
for instance, 
how will this difference evolve?
At the end of the coarsening period, each enantiomer shows a 
cluster size distribution similar to that shown by a box graph in Fig. 8(a).
The number of small clusters $s<s_c$ with large negative velocities $v(s)$ 
vanish except the number of monomers. 
Since monomers changed their chirality often enough,
their numbers turn out to be close for both enantiomers: 
$C_R(1) \approx C_S(1)$.
The numbers of clusters of sizes 4 and 6 remain large, which is due to the smallness 
of their velocities in size space.
Their contribution to the summation on the right-hand side of eq.(\ref{eq4.30})
is negligible.
Thus, the amplification rate of number difference $d M/dt$ is governed by
the contribution from the cluster size region with a positive velocity $v(s)$.
Then, the rate $d M/dt$
 is positive if $C_R(s)>C_S(s)$ for $s>s_c$.
By assuming that the dominant contribution originates from those terms around 
the average cluster size $\bar s  $, the velocity $v(s)$ is replaced by 
$v(\bar s  )>0$, 
and then, the right-hand side of eq.(\ref{eq4.30})
is transformed as
\begin{align}
\frac{d M}{d t}= v(\bar s  ) (C_R- C_S)|_{t=(i+1)/f} = \frac{v(\bar s)}{\bar s}
M^{(i+1)},
\label{eq4.32}
\end{align}
where $C_R$ and $C_S$ are  the total numbers of clusters, defined in 
eq.(\ref{eq4.24}) and evaluated just before the $(i+1)$th grinding.
In the last equality, the approximation $\bar s_R=\bar s_S=\bar s$ is used,
 and $M^{(i+1)}=\bar s (C_R-C_S)|_{t=(i+1)/f}$ 
 is the difference in number between the R and S enantiomeric molecules
before the $(i+1)$th grinding.

By assuming that 
the rate $dM/dt$ remains to be of the order given by eq.(\ref{eq4.32})
in almost the entire coarsening period between the consecutive
$i$th and $(i+1)$th  grinding processes, 
then by integration over a period $1/f$, one obtains
\begin{align}
M^{(i+1)}- M^{(i)}= 
\frac{v(\bar s)}{f \bar s} M^{(i+1)}.
\end{align}
Note that grinding does not affect the total numbers of enantiomeric molecules
$N_R$ and $N_S$, and thus, the value of $M$ after the $i$th grinding is the same as
that before the grinding $M^{(i)}$.
Since the chiral order parameter is written in terms of the number difference
$M$ as $\phi=M/N$,
it increases exponentially after $i$th grinding at the time $t=i/f$ as
\begin{align}
\phi^{(i)}= \Big( 1-\frac{v(\bar s)}{\bar s f} \Big)^{-i} \phi^{(0)}
= \phi^{(0)} \exp (t/\tau)
\end{align}
with 
\begin{align}
\tau=-\frac{1}{f \log \Big( 1-\frac{v(\bar s)}{\bar s f} \Big)}
\approx \frac{\bar s}{ v(\bar s)}.
\label{eq4.36}
\end{align}
Thus, the homochirality $|\phi|=1$ should be accomplished 
at $t_s=-\tau \ln \phi^{(0)}$, irrelevant of the system size,
in agreement with the results shown in Fig. 4. 

On simply decreasing the grinding frequency from $f=1$ to $f=0.1$,
the average cluster size increases from $ \bar s=18$, as shown in Fig. 8,
 to $\bar s \approx 24$ and
the saturation time 
varies
from $t_s\nu_0=8833$ to $t_s\nu_0=15173$:
The larger the average cluster size $\bar s$, the longer the saturation time $t_s$,
not in contradiction to eq. (\ref{eq4.36}).
However, the precise behavior of $t_s$ requires knowledge on $v(\bar s)$ and
other factors, which have been difficult to evaluate thus far.

\subsection{With grinding and an enhancement of chirality conversion 
rate
}

In addition to grinding, we can incorporate the enhancement of chirality conversion
rate
when an isolated enantiomer is in contact with clusters of opposite chirality.
For example, the conversion rate is enhanced to 
$\nu_n=10^n\nu_0$ when a monomer is in contact with $n$ opposite enantiomers.
$n=1$ indicates that the monomer is 
on the flat face
 of the other cluster, and $n=2$
at the kink site.
The enhancement in chirality conversion actually speeds up the establishment 
of homochirality,
as is depicted in the plot of saturation time $t_s$ by up-triangles in Fig. 4.
The saturation time $t_s$ is by far shorter than that in the case without enhancement.
If, for instance, we include such
 enhancement in our sample system with $L^2=80^2$
at $\nu_0/f=1$, the saturation time shortens to $t_s \nu_0=590$, instead
of 8834 without enhancement. 

In our previous study, we have simulated only for a short time
and thus concluded that the enhancement is essential for achieving homochirality.
 Results of the present analysis, however, 
clearly indicates that the enhancement speeds up the process
but does not affect essential features qualitatively.
The grinding alone is sufficient for establishing homochirality
within a saturation time, which is independent of
the system size. 
In fact,
when the variations in average cluster size $\bar s_{R,S}$ and
population $C_{R,S}$ are plotted with the scaled time $t/t_s$, 
a universal feature reveals itself. Symbols of average sizes $\bar s_{R,S}$ 
(Fig. 9(a)) and cluster populations $C_{R,S}$ (Fig. 9b) 
collapse on those obtained without
enhancement, shown by curves, in the previous section.

\begin{figure}[tbh]
\begin{center} 
\begin{minipage}{0.48\linewidth}
\begin{center}
\includegraphics[width=1.0\linewidth]{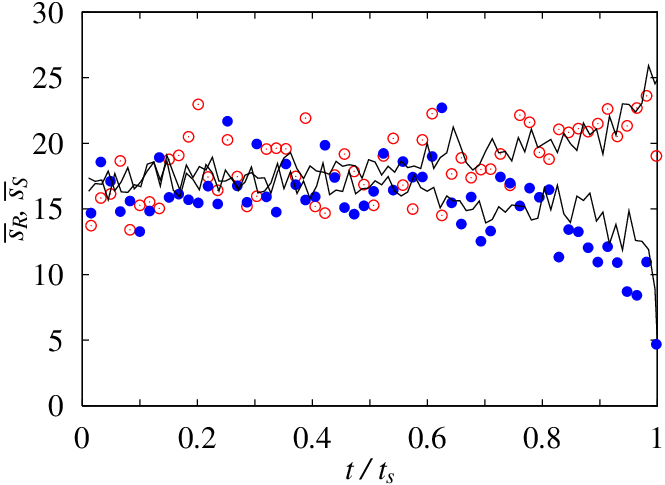}\\
(a)\\
\end{center}
\end{minipage}
\begin{minipage}{0.48\linewidth}
\begin{center}
\includegraphics[width=1.0\linewidth]{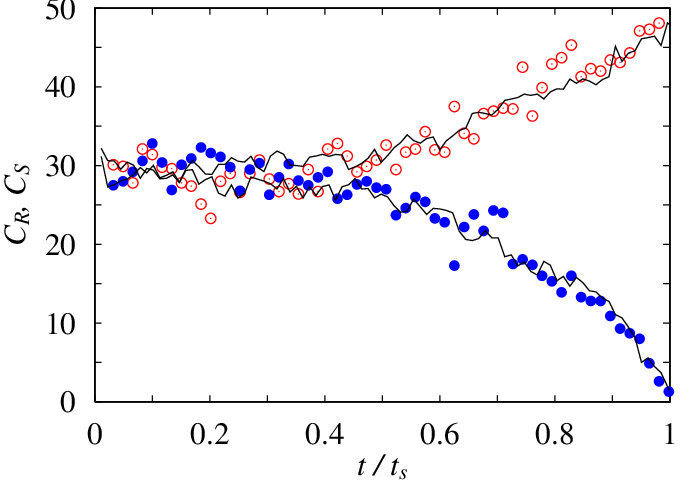}\\
(b)\\
\end{center}
\end{minipage}
\end{center} 
\caption{(Color online) Scaling plot of temporal evolution of
(a) average sizes $\bar s_{R,S}$ and 
(b) cluster populations $C_{R,S}$ of R and S clusters
 under grinding with (symbols) and without
(curves) enhancement of chirality conversion rate on the cluster of opposite
chirality.
The system size is $L^2=80^2$ at a number of molecules $N=640$ 
at a concentration $c=0.1$. The grinding frequency is $\nu_0/f=1$.
}
\label{fig9}
\end{figure}

\section{Conclusion}

A lattice gas model for conglomerate growth is shown to be 
reduced to a spin-one Ising model, the so-called Blume-Emery-Griffiths model
\cite{blume+71}.
Then, chiral symmetry breaking is undestood as 
an equilibrium phase transition.
However, the establishment of the true equilibrium state with homochirality
is hampered by the very slow process of Ostwald ripening:
Two large clusters of different enantiomers compete with each other, but
the driving force for chirality selection reduces its strength
as the system size increases.
In our simple two-dimensional model 
where crystal growth is controlled by surface kinetics,
the time necessary to achieve homochirality is proportional to the
total number of molecules. 
In a macroscopic system, one cannot effectively 
reach the homochiral state with a single large crystal cluster \cite{noorduin+08b}.
If diffusion comes into play instead of stirring, 
the necessary time should be longer.
In actual experiments on solution growth, the system usually ends up with many 
crystal clusters with different chiralities.\cite{kipping+98,kondepudi+90,noorduin+08b}

The grinding process forcibly breaks crystal clusters, and 
cluster size is distributed in a certain range.
The width of the range and the average size of clusters are almost independent of
the system size, since they are essentially determined 
by the grinding frequency and mesh size.
To compensate for the size restriction, the number of clusters becomes
 proportional to the system size. 
Under perpetual grinding, monomers and small clusters are constantly
supplied by recycling and high supersaturation is 
maintained,
 since the system is closed.
It allows the growth rate of large clusters to stay positive.
The positive growth rate in size space amplifies 
the difference in total cluster number between two enantiomers,
and the difference increases exponentially with time.
The saturation time required to establish the
homochiral state depends on the average cluster size, but
it is independent of the system size.
The result is consistent with that of experiments with grinding where
homochiral state is achieved within an observable time.
\cite{viedma05,noorduin+08a,noorduin+08b}

The enhancement of chirality conversion of monomers
at the cluster surface shortens the saturation time, 
but the essential feature of chirality selection does not require enhancement,
contrary to the conclusion in our previous paper \cite{saito+08}.

In a recent paper by Tsogoeva and coworkers\cite{tsogoeva+09}, 
conglomerate crystallization of chiral organic molecule is
studied. The molecule changes its chirality through a Mannich type
reversible reaction.
By assuming that the reaction proceeds fast enough so that a local equilibrium
between achiral reactants and chiral products is established, we can eliminate the 
degrees of freedom of reactants. 
Then, chirality conversion reduces to the reaction process Eq.(\ref{eq3.15}),
and our analysis possibly applies to this case.
For general cases, further study is necessary.

Grinding the system means that one has to put energy into the system from outside
to break bondings between molecules.
The system is closed but driven energetically.
Therefore, even though the conglomerating interaction assures chiral
symmetry breaking in thermal equilibrium, the eventual homochiral state 
is far away from the equilibrium state:
Final steady state consists of many small clusters, 
instead of a single large crystal corresponding to the thermal equilibrium state.
Is there any relevance of the symmetry breaking in crystal chirality to
the homochirality in life,
notwithstanding that the equilibrium state of the chiral crystal growth is 
homochiral at low temperatures while that of the chemical reaction system 
in life may be racemic?  
The present work shows that the homochiral state in crystal growth is 
actually a nonequilibrium state achieved and sustained by external driving.
This external driving together with the closedness of the system enhances the nonlinear production due to supersaturation and recycling. This scenario 
corresponds to that proposed in ref. 6
 in a chemical reaction system. 
Thus, to attain homochirality in a chemical system, it seems necessary
to place it under external driving, or in other words, homochirality in life 
may be possible in a dissipative state.

{\bf Acknowledgement}\\
Y.S. acknowledges support by a Grant-in-Aid for Scientific Research 
(No. 19540410) from the Japan Society for the
Promotion of Science.


\end{document}